\documentclass[graybox]{svmult}

\usepackage{type1cm}        
\usepackage{makeidx}         
\usepackage{graphicx}        
\usepackage{multicol}        
\usepackage[bottom]{footmisc}
\usepackage{rotating}
\usepackage{array}
\usepackage{hyperref}
\usepackage{comment}
\usepackage{orcidlink}

\usepackage{newtxtext}       %
\usepackage[varvw]{newtxmath}       

\makeindex             

\begin{document}

\title*{Deep Generative Models for 3D Medical Image Synthesis}
\author{Paul Friedrich\\
Yannik Frisch\\ 
Philippe C. Cattin}
\authorrunning{P. Friedrich et al.}
\institute{Paul Friedrich \orcidlink{0000-0003-3653-5624} \at Department of Biomedical Engineering, University of Basel, Hegenheimermattweg 167b, 4123 Allschwil, Switzerland \email{paul.friedrich@unibas.ch}
\and Yannik Frisch \orcidlink{0009-0005-8097-0158} \at Graphical-Interactive Systems, Technical University Darmstadt, Fraunhoferstr. 5, 64283 Darmstadt, Germany \email{yannik.frisch@gris.tu-darmstadt.de}
\and Philippe C. Cattin \orcidlink{0000-0001-8785-2713} \at Department of Biomedical Engineering, University of Basel, Hegenheimermattweg 167b, 4123 Allschwil, Switzerland \email{philippe.cattin@unibas.ch}}
%
%
\maketitle


\abstract{Deep generative modeling has emerged as a powerful tool for synthesizing realistic medical images, driving advances in medical image analysis, disease diagnosis, and treatment planning. This chapter explores various deep generative models for 3D medical image synthesis, with a focus on Variational Autoencoders (VAEs), Generative Adversarial Networks (GANs), and Denoising Diffusion Models (DDMs). We discuss the fundamental principles, recent advances, as well as strengths and weaknesses of these models and examine their applications in clinically relevant problems, including unconditional and conditional generation tasks like image-to-image translation and image reconstruction. We additionally review commonly used evaluation metrics for assessing image fidelity, diversity, utility, and privacy and provide an overview of current challenges in the field.}

\section{Introduction}
\label{sec:introduction}
Medical imaging plays a critical role in diagnosing, monitoring, and treating disease by providing unique structural, functional, and metabolic information about the human body. While natural images usually capture data in two dimensions, medical practice often requires the acquisition of three-dimensional volumes like Magnetic Resonance Imaging (MRI), Computed Tomography (CT), or Positron Emission Tomography (PET) scans. Acquiring these volumetric scans can be time-consuming, costly, limited by scanner availability, and in the case of CT and PET scans, expose patients to harmful radiation. In addition, privacy and ethical concerns make it difficult to share medical data. Together, these factors limit the availability of large-scale medical image datasets for scientific studies, deep learning in medical image analysis, or physician training. 

\begin{figure}
    \centering
    \includegraphics[width=.8\textwidth]{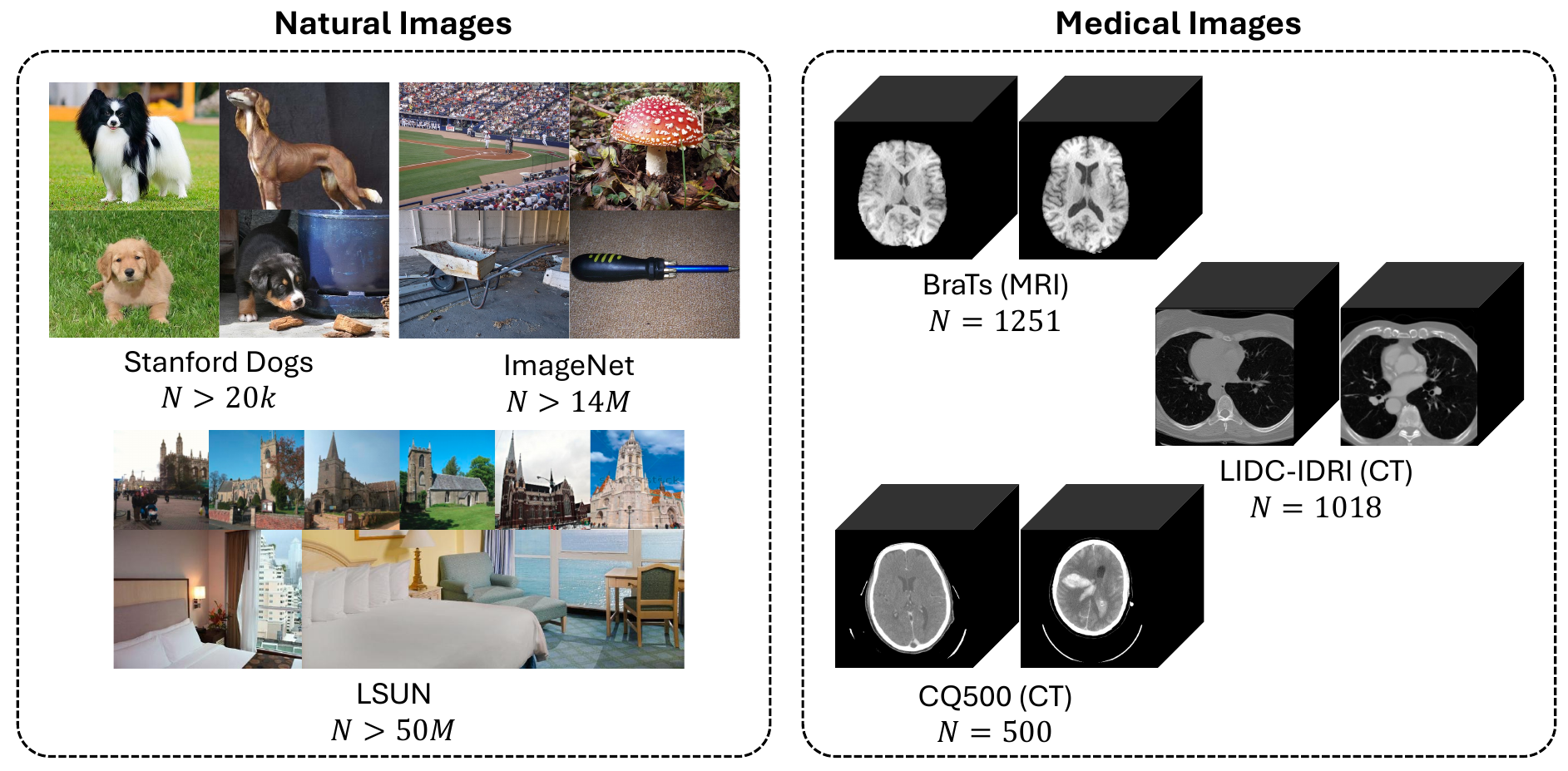}
    \caption{Differences between natural and medical images. Natural 2D images \cite{deng2009imagenet,khosla2011dogs,yu2015lsun} are widely available in large-scale datasets, as they can easily be scraped from the internet. In contrast, 3D medical data \cite{bakas2017advancing,bakas2018identifying,chilamkurthy2018development,armato2011lung,menze2014multimodal} is scarce due to the high cost of acquisition, as well as ethical and privacy concerns.}
    \label{fig:diff_nat_med}
\end{figure}

Driven by advances in generating synthetic natural images, the application of deep generative models to medical images has emerged as a promising solution to address data scarcity and enable various medical image analysis tasks \cite{fernandez20233d,friedrich2024wdm,frisch2023synthesising,konz2024anatomically}. However, the three-dimensionality and distinct distribution characteristics of medical images, shown in Figure~\ref{fig:diff_nat_med}, present unique challenges for image synthesis, requiring a careful adaptation of standard methods \cite{wen2021rethinking}. This chapter explores the basics of popular image generation models such as Variational Autoencoders (VAEs), Generative Adversarial Networks (GANs), and Denoising Diffusion Models (DDMs), discusses the advantages and disadvantages of each model type, reviews their applications in various medical imaging tasks, and takes a look at common evaluation metrics for assessing model performance.
\subsection{Deep Generative Models}
\label{subsec:DGM}
\begin{figure}
    \centering
    \includegraphics[width=.9\textwidth]{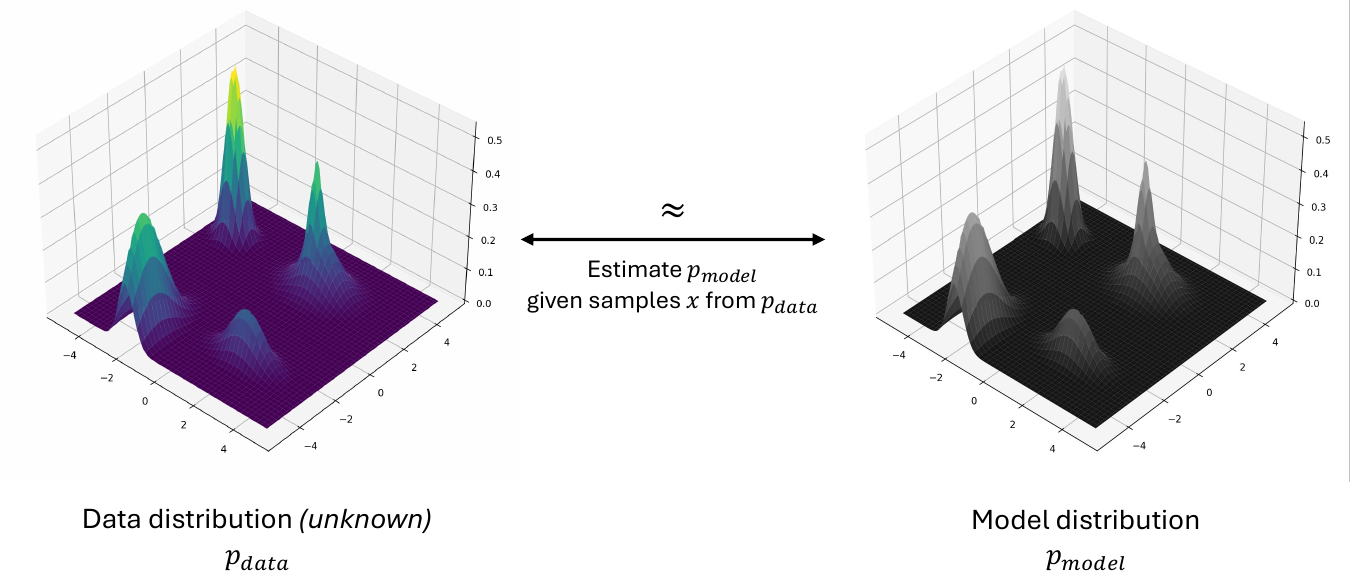}
    \caption{The basic principle of generative modeling. Using data from the data distribution $p_{data}$, we try to find a model $p_{model}$ that closely follows this distribution. We can then use this model to generate new samples that resemble the original data distribution.}
    \label{fig:dgm_overview}
\end{figure}

Generative models are a class of machine learning models that aim to learn the underlying distribution $p_{data}$ of some input data $x$ to (1) generate new samples from that same distribution or (2) assign probability values to existing samples, allowing for certain downstream tasks. In Deep Generative Models (DGMs), this probability density estimation task is solved using deep neural networks that either explicitly model the distribution $p_{model}$ or parameterize a model that can sample from $p_{model}$ without explicitly estimating it. This general principle of finding a model that accurately represents the underlying data distribution of some input data $x$ is shown in Figure~\ref{fig:dgm_overview}.

In recent years, deep generative modeling has been applied to various data modalities, including text \cite{vaswani2017attention}, audio \cite{oord2016wavenet}, shapes \cite{luo2021diffusion}, and images \cite{goodfellow2014generative}, using models such as Restricted Boltzmann Machines \cite{freund1991unsupervised}, Normalizing Flows \cite{rezende2015variational}, Variational Autoencoders \cite{kingma2013auto}, Generative Adversarial Networks \cite{goodfellow2014generative}, and Denoising Diffusion Models \cite{ho2020denoising,sohl2015deep}.
\section{Background on Deep Generative Models}
\label{sec:background_dgm}
The following section focuses on deep generative modeling of 3D medical images $x \in \mathbb{R}^{D \times H \times W}$ with VAEs, GANs and DDMs, as these models are widely used in medical image computing. We will briefly describe the basic principle of each model, provide insights into their training objectives, and discuss the advantages and disadvantages of the different networks. 
\subsection{Variational Autoencoders}
\label{subsec:backgroundVAE}
\begin{figure}
    \centering
    \includegraphics[width=.8\textwidth]{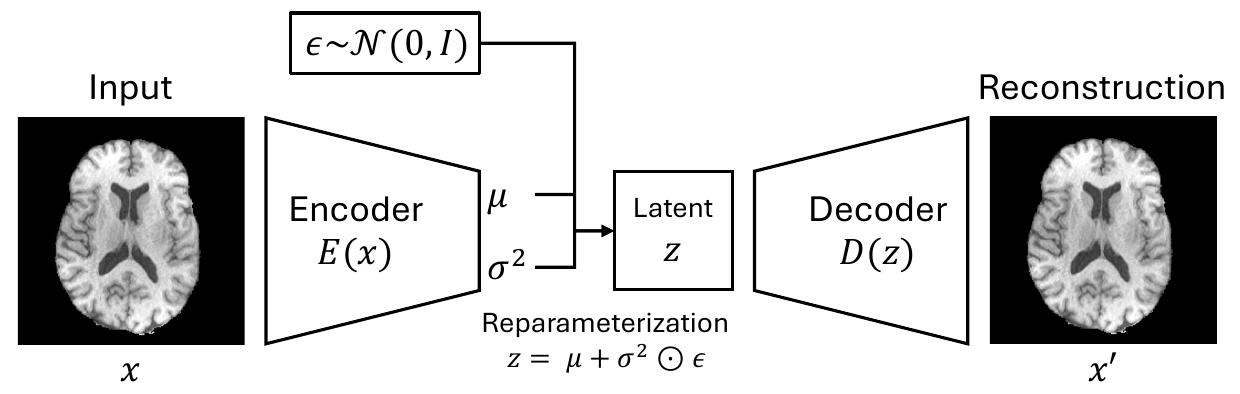}
    \caption{The basic principle of Variational Autoencoders. An input image $x$ is encoded into a KL regularized latent representation $z = E(x)$ and is subsequently reconstructed as $x'=D(z)$. By minimizing the reconstruction error, as well as the KL-divergence between the latent and a standard normal distribution, the model learns to generate new data and encode data in a meaningful way.}
    \label{fig:vae_overview}
\end{figure}

The basic principle of VAEs \cite{kingma2013auto,rezende2014stochastic} builds on that of standard autoencoders. Both encode an input $x$, e.g. a 3D medical image, into a low-dimensional latent representation $z=E(x)$ using an encoder network $E$. 
The original image $x'=D(z)$ is then reconstructed from this representation $z$ using a decoder network $D$. VAEs, however, differ in the way they parameterize this latent representation. Instead of directly encoding the image $x$ into a single vector $z$, they encode it into the parameters of a normal distribution by designing the encoder in a way to predict the mean $\mu= E_\mu(x)$ and variance $\sigma^2 =E_\sigma(x)$ of that distribution. The latent representation $z$ can then be drawn from $\mathcal{N}(\mu, \sigma^2)$. As backpropagating through this stochastic part would be impossible, VAEs apply a reparameterization trick and instead sample an auxiliary variable $\epsilon\sim\mathcal{N}(0,I)$ to define $z=\mu + \sigma^2 \odot \epsilon$. In addition, VAEs apply a KL-divergence regularization term to this latent distribution to make it close to a standard normal distribution. This allows generating new samples by drawing $z\sim\mathcal{N}(0,I)$ and passing it through the trained decoder network. This general setup is shown in Figure~\ref{fig:vae_overview}.
Combining these two principles, VAEs can be trained by minimizing the reconstruction error between input and reconstructed image, as well as by applying the KL-divergence regularization term to the latent representation, which results in the following training objective:
\begin{equation}
    \mathcal{L}_{VAE} = \|x - D(E(x))\|_2^2 - D_{KL}(\mathcal{N}(E_\mu(x), E_\sigma(x))\|\mathcal{N}(0,I)).
\end{equation}
This objective maximizes the evidence lower bound (ELBO) on the log-likelihood of the data. The model, therefore, learns to generate new data and compress data into a meaningful latent representation.
While VAEs have a relatively short inference time and are known for their ability to produce diverse images, they suffer from poor sample quality and often produce blurry images \cite{xiao2021tackling}. To overcome this problem, variations such as Vector Quantized VAEs (VQ-VAEs) \cite{van2017neural} were proposed that map to a discrete learned instead of a continuous static latent distribution. Another commonly used variant is VQ-GAN \cite{esser2021taming}, which combines the VQ-VAE concept with the adversarial training of GANs, another DGM that we will discuss in the next section.

\subsection{Generative Adversarial Networks}
\label{subsec:backgroundGAN}
In recent years, GANs \cite{goodfellow2014generative} have successfully been used to generate medical images, and form the basis of many applications in the medical field. Unlike most other generative models, GANs don't explicitly model the underlying data distribution in terms of a probability density function but take a different approach by implicitly modeling the distribution through a process of adversarial training.
\begin{figure}
    \centering
    \includegraphics[width=.9\textwidth]{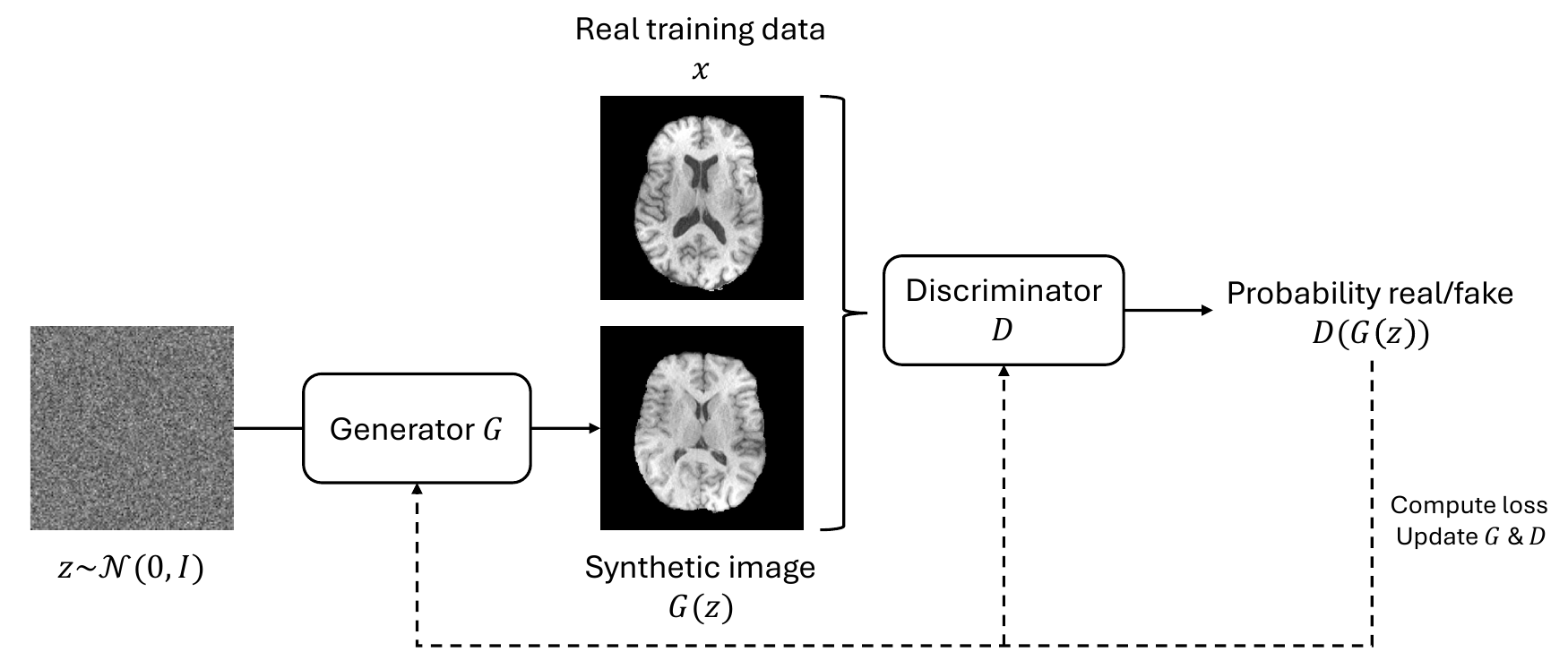}
    \caption{The basic principle of Generative Adversarial Networks. The generator $G$ and the discriminator $D$ play an adversarial game against each other, where the generator tries to synthesize realistic images that the discriminator cannot distinguish from the real training data.}
    \label{fig:gan_overview}
\end{figure}
GANs are trained following a two-player min-max game shown in Figure~\ref{fig:gan_overview}, and generally consist of two networks: the Generator $G(z)$ that aims to generate realistic fake samples from random noise $z \sim \mathcal{N}(0,I)$ and the Discriminator $D(x)$ that tries to distinguish real and fake samples by solving a classification task. Both networks are iteratively optimized using the following training objective:
\begin{equation}
    \min_{G}\max_{D} \mathcal{L}_{GAN}(D,G) = \mathbb{E}_{x\sim p_{data}}[\log D(x)] + \mathbb{E}_{z\sim \mathcal{N}(0,I)}[\log(1-D(G(z)))]
\end{equation}
Intuitively, the generator aims to produce increasingly realistic fake data to fool the discriminator, while the discriminator tries to get better at distinguishing real data from fake data.
While GANs have demonstrated impressive image generation capabilities \cite{brock2018large,karras2020analyzing}, they often suffer from problems such as training instabilities and convergence problems. These issues can be caused by a mismatch between the capacity of the generator and the discriminator or an overconfident discriminator that makes it difficult for the generator to learn and optimize its parameters. Another common problem is mode collapse, where the generator learns to output limited variations of samples by ignoring certain modes of the data distribution. To overcome these problems, various GAN modifications that apply improved training techniques, regularization strategies, or loss modifications have been introduced. Those include Wasserstein GAN (WGAN)\cite{arjovsky2017wasserstein}, WGAN with Gradient Penalty (WGAN-GP) \cite{gulrajani2017improved}, Spectral Normalization GAN (SNGAN) \cite{miyato2018spectral}, or Least Squares GAN (LSGAN) \cite{mao2017least}. These adaptations have successfully reduced GAN-related problems but do not completely eliminate them.

\subsection{Denoising Diffusion Models}
\label{subsec:background_ddpm}
\begin{figure}
    \centering
    \includegraphics[width=.9\textwidth]{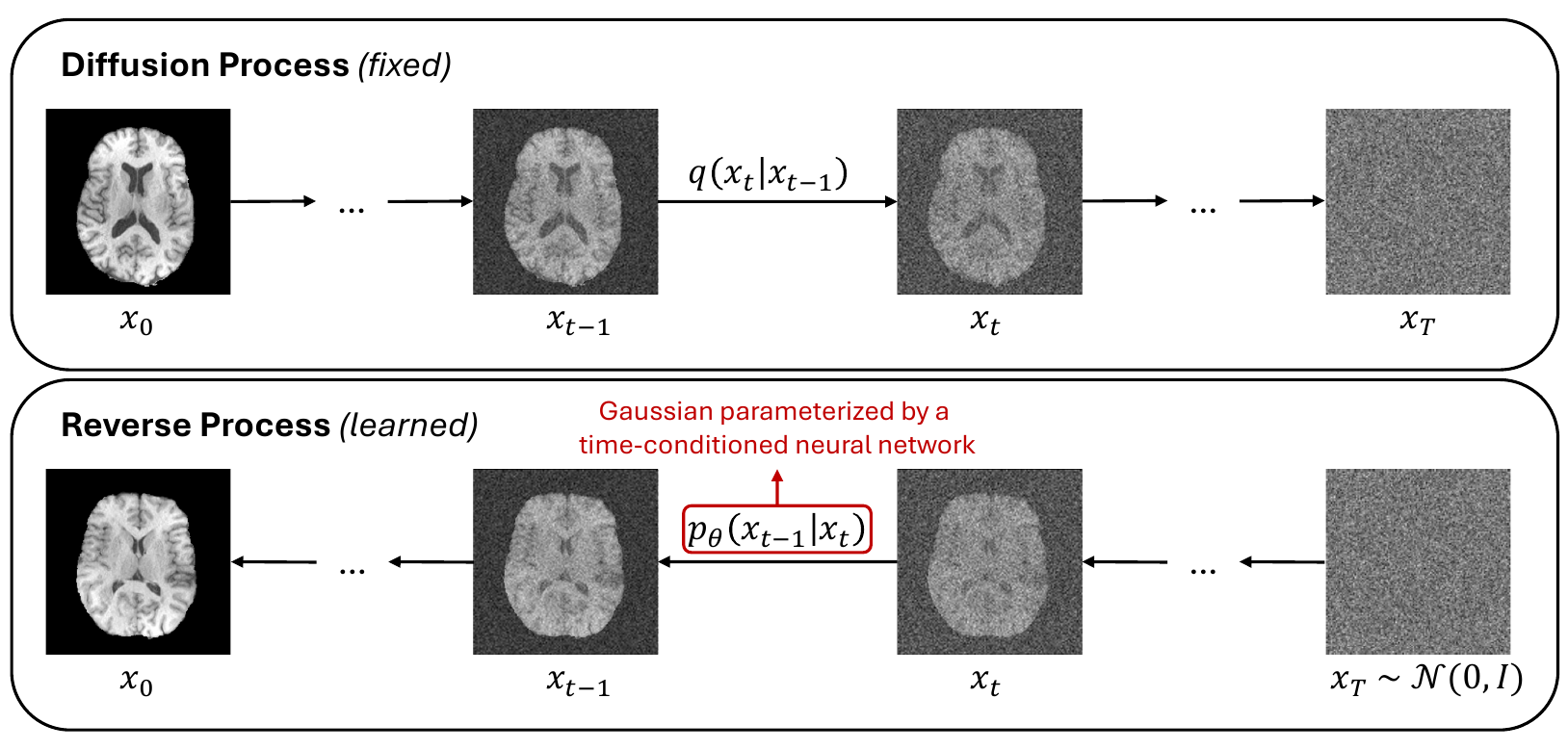}
    \caption{The basic principle of Denoising Diffusion Models. The diffusion model consists of two main components: a fixed \textit{diffusion process} that gradually perturbs input data with Gaussian noise and maps the data distribution to a simple prior, and a learned \textit{reverse process} with each transition being a Gaussian parameterized by a time-conditioned neural network.}
    \label{fig:ddm_overview}
\end{figure}

Denoising Diffusion Models \cite{ho2020denoising,sohl2015deep} are latent variable models that sample from a distribution by reversing a defined diffusion process. This \textit{diffusion} or \textit{forward process} progressively perturbs the input data with Gaussian noise and maps the data distribution to a simple prior, namely a standard normal distribution. To generate new samples, we aim to learn the \textit{reverse process}, which maps from this prior to the data distribution. New samples are generated by drawing random noise from the prior and passing it through the reverse process. This general principle is shown in Figure~\ref{fig:ddm_overview}. 
The diffusion process consists of $T$ timesteps and can be described as a Markov chain, with each transition being a Gaussian that follows a predefined variance schedule $\beta_{1}, ..., \beta_{T}$:
\begin{equation}
    q(x_{1:T}|x_{0}) := \prod_{t=1}^{T}q(x_{t}|x_{t-1}), \quad \text{with} \quad q(x_{t}|x_{t-1}):= \mathcal{N}(\sqrt{1-\beta_{t}}x_{t-1}, \beta_{t}I).
\end{equation}
The reverse process can also be described as a Markov chain with learned Gaussian transition kernels, starting at a simple prior distribution $p(x_T)=\mathcal{N}(0,I)$:
\begin{equation}
    p_{\theta}(x_{0:T}) := \prod_{t=1}^{T}p_{\theta}(x_{t-1}|x_{t}), \quad \text{with} \quad p_{\theta}(x_{t-1}|x_{t}):= \mathcal{N}(\mu_{\theta}(x_t, t), \Sigma_{\theta}(x_t, t)).
\end{equation}
While $\Sigma_{\theta}(x_t, t)$ is often fixed to the forward process variances $\beta_t$, $\mu_{\theta}(x_t, t)$ is usually estimated by a time-conditioned neural network. This network is trained by minimizing the variational lower bound of the negative log-likelihood. Following a reparameterization trick \cite{ho2020denoising}, we can configure the network to predict the noise $\epsilon_\theta(x_t, t)$ to be removed from a corrupted sample $x_t$ and simplify the training objective to an MSE loss, with $\alpha_t = 1 - \beta_t$, $\bar{\alpha}_t = \prod_{s=1}^t\alpha_s$, and $\epsilon \sim \mathcal{N}(0, \boldsymbol{I})$:
\begin{equation}
    \mathcal{L}_{simple} =  \| \epsilon - \epsilon_\theta(x_t, t) \|^2_2, \quad \text{where} \quad x_t = \sqrt{\bar{\alpha}_t} x_0 + \sqrt{1-\bar{\alpha}_t}\epsilon.
\end{equation}
Given a trained network $\epsilon_\theta(x_t, t)$ and a randomly drawn starting point $x_T\sim\mathcal{N}(0,I)$, we can iteratively produce a new sample by applying the following equation $T$ times:
\begin{equation}
    x_{t-1} = \frac{1}{\sqrt{\alpha_t}}\left(x_t - \frac{1-\alpha_t}{\sqrt{1-\bar{\alpha}_t}} \epsilon_\theta(x_t, t)\right) + \sigma_t\epsilon
\end{equation}
While diffusion models have demonstrated impressive image generation capabilities \cite{dhariwal2021diffusion}, their iterative nature requires multiple network evaluations for generating a single sample, making them slow and resource-intensive. To speed up this sampling process, several adaptations have been introduced, such as the Denosing Diffusion Implicit Model (DDIM) \cite{song2020denoising}, which formulates a deterministic non-Markovian process to sample with fewer steps, different knowledge distillation methods, such as consistency distillation \cite{song2023consistency}, or combinations of different approaches, such as adversarial training of the denoising network, as proposed in DDGAN \cite{xiao2021tackling}.
\subsubsection{Latent Diffusion Models}
\label{subsubsec:LDM}
\begin{figure}
    \centering
    \includegraphics[width=.9\textwidth]{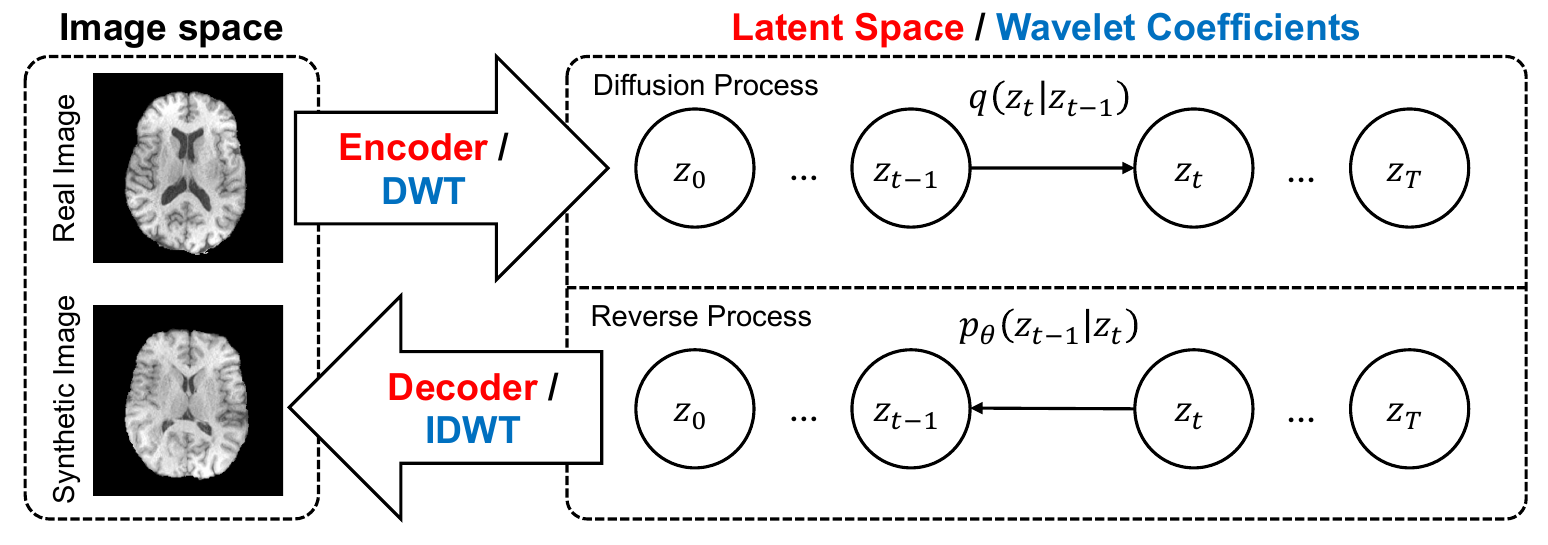}
    \caption{The basic principle of Latent Diffusion Models (red) and Wavelet Diffusion Models (blue).}
    \label{fig:ldm_wdm_overview}
\end{figure}

To reduce the computational complexity of standard DDMs, Latent Diffusion Models (LDMs) \cite{rombach2022high} have been introduced. While LDMs share the same fundamental principle as standard denoising diffusion models, they operate on a learned, more compact latent representation of the data rather than directly on the images. Training LDMs begins with training an autoencoder, such as VQ-GAN \cite{esser2021taming}, to generate a meaningful low-dimensional latent representation of the data. Subsequently, the diffusion model is trained on this latent representation $z$ instead of the original high-dimensional data $x$, resulting in a more computationally efficient approach. This principle is illustrated in Figure~\ref{fig:ldm_wdm_overview}. 
To generate new samples, the reverse diffusion process is applied starting from random noise $z_T\sim\mathcal{N}(0,I)$ in the latent space, producing a synthetic latent representation $z_0$. This latent representation is then decoded back into the image space using the trained decoder. Although LDMs effectively reduce the computational complexity of training and sampling from denoising diffusion models, they depend on a well-performing autoencoder. Training such an autoencoder for high-resolution medical volumes is challenging itself and often constrained by computational resources. 
\subsubsection{Wavelet Diffusion Models}
\label{subsubsec:WDM}
Wavelet diffusion models (WDMs) \cite{friedrich2024wdm,phung2023wavelet}, shown in Figure~\ref{fig:ldm_wdm_overview}, are a promising alternative to LDMs. While both approaches share a similar idea, wavelet diffusion models take a different approach to spatial dimensionality reduction by applying Discrete Wavelet Transform (DWT). This approach is learning-free in the sense that it does not require a pre-trained autoencoder. The diffusion model then operates on the wavelet coefficients $z$ of the images. To generate new images, WDMs start with random noise $z_T\sim\mathcal{N}(0,I)$ and apply the reverse diffusion process to produce synthetic wavelet coefficients $z_0$. These coefficients can then be transformed back to the image space using Inverse Discrete Wavelet Transform (IDWT). As WDMs do not rely on an autoencoder network, they have an even smaller memory footprint than LDMs, which is particularly important in 3D medical image synthesis tasks that are typically constrained by the available GPU memory.

\section{Applications in Medical Image Computing}
The following section provides an overview of DGMs in 3D medical image synthesis. Starting with the fundamental task of unconditional image generation, followed by a discussion of conditional image generation tasks, including image-to-image translation and image reconstruction. It is important to note that this only covers a subset of potential applications of DGMs in medicine. Additional tasks, such as image registration \cite{kim2022diffusemorph}, classification \cite{li2023your}, segmentation \cite{wolleb2022seg,wu2024medsegdiff}, anomaly detection \cite{sanchez2022healthy,wolleb2022diffusion}, and image inpainting \cite{durrer2024adenoising,friedrich2023point} can also be addressed using generative networks. 

\subsection{Unconditional Image Generation}
Unconditional image generation involves the synthesis of new images without any specific condition and simply demonstrates the ability of a generative model to learn the underlying distribution of some given training data. These models can be applied to augment datasets with synthetic images \cite{ye2023synthetic} or improve downstream applications fairness under distribution shifts \cite{ktena2024generative}. Unconditional image generation is commonly used to demonstrate the performance of novel architectures, which serve as a foundation for developing methods for conditional generation tasks. An overview of publications that present unconditional 3D medical image synthesis models is given in Table~\ref{table:overview_uncond}, where we provide information on the synthesized image modality, the image resolution, the public datasets used for training, as well as the network architecture.
\begin{table}
\begin{center}
\caption{Overview of publications on unconditional 3D medical image synthesis models, the modality of the generated data, public datasets used for training, the maximum image resolution reported in the paper, as well as the utilized network architecture.}
\label{table:overview_uncond}
\resizebox{\textwidth}{!}{
\begin{tabular}{ |l|c|c|p{1.5cm}|c|c|c|c| } 
 \hline
 \textbf{Study} & \textbf{Model} & \textbf{Modality} & \textbf{Dataset(s)} & \textbf{Resolution} & \textbf{Network(s)}\\ 
 \hline\hline
 Volokitin et al. (2020) \cite{volokitin2020modelling} & VAE & MRI & HCP & $256 \times 256 \times 256$ & 2D Slice-VAEs \\
 Kapoor et al. (2023) \cite{kapoor2023multiscale} & VAE & MRI & ADNI & $80 \times 96 \times 80$ & 3D VAE \\
 \hline
 Kwon et al. (2019) \cite{kwon2019generation} & GAN & MRI & ADNI, BRATS, ATLAS & $64 \times 64 \times 64$ & 3D $\alpha$-WGAN-GP \\
 Segato et al. (2020) \cite{segato2020data} & GAN & MRI & ADNI & $64 \times 64 \times 64$ & 3D $\alpha$-GAN \\
 Granstedt et al. (2021) \cite{granstedt2021slabgan} & GAN & MRI & fastMRI & $256 \times 256 \times 16$ & 3D GAN \\
 Hong et al. (2021) \cite{hong20213d} & GAN & MRI & ADNI, ABIDE, ADHD2000, HABS, GSP, MCIC, OASIS, PPMI & $80 \times 96 \times 112$ & 3D StyleGAN \\
 Chong et al. (2021) \cite{chong2021synthesis} & GAN & MRI & HCP & $256 \times 256 \times 256$ & 3D WGAN-GP + 2D pix2pixGAN \\
 Bergen et al. (2022) \cite{bergen20223d} & GAN & PET & HECKTOR & $64 \times 64 \times 32$ & 2D TGAN \\
 Mensing et al. (2022) \cite{mensing20223d} & GAN & MRI & German National Cohort &$160 \times 160 \times 128$ & 3D cGAN (FastGAN)\\
 Sun et al. (2022) \cite{sun2022hierarchical} & GAN & MRI, CT & GSP, COPDGene & $256 \times 256 \times 256$ & 3D HA-GAN \\
 Liu et al. (2023) \cite{liu2023inflating} & GAN & MRI & COCA, ADNI & $64 \times 64 \times 64$ & 2D StyleGAN2 + 3D StyleGAN2 \\
 Kim et al. (2024) \cite{kim20243d} & GAN & MRI, CT & HCP, CT-ORG & $128 \times 128 \times 128$ & 3D WGAN-GP \\
 \hline
 Dorjsembe et al. (2022) \cite{dorjsembe2022three} & DDM & MRI & ICTS & $128 \times 128 \times 128$ & 3D DDPM \\ 
 Pinaya et al. (2022) \cite{pinaya2022brain} & LDM & MRI & UK Biobank & $160 \times 224 \times 160$ & 3D VAE + 3D DDPM/DDIM \\
 Peng et al. (2023) \cite{peng2023generating} & DDM & MRI & ADNI, UCSF, SRI International & $128 \times 128 \times 128$ & 2.5D DDPM \\
 Khader et al. (2023) \cite{khader2023denoising} & LDM & MRI, CT & DUKE, MRNet, ADNI, LIDC-IDRI & $128 \times 128 \times 128$ & 3D VQ-GAN + 3D DDPM \\
 Friedrich et al. (2024) \cite{friedrich2024wdm} & WDM & MRI, CT & BRATS, LIDC-IDRI & $256 \times 256 \times 256$ & 3D WDM (DDPM)\\
 \hline
 \end{tabular}}
\end{center}
\end{table}

Although VAEs are widely used for modeling 2D medical images, their application to unconditional 3D image generation remains relatively unexplored. Volokitin et al. \cite{volokitin2020modelling} used 2D slice VAEs to model high-resolution 3D brain MR images by combining a VAE with a Gaussian model that allows for sampling coherent stacks of latent codes that decode into a meaningful volume. Kapoor et al. \cite{kapoor2023multiscale} took a different approach by transforming a reference MRI with multi-scale morphological transformations predicted by a 3D VAE.

GAN-based approaches have been extensively explored for 3D medical image synthesis, with early applications successfully modeling brain MR images. Kwon et al. \cite{kwon2019generation} and Segato et al. \cite{segato2020data} adapted the $\alpha$-GAN framework \cite{lutz2018alphagan} for this purpose, while Hong et al. \cite{hong20213d} presented a 3D version of StyleGAN \cite{karras2020analyzing}. While these early approaches were constrained to relatively low resolutions, Chong et al. \cite{chong2021synthesis} were the first to scale GAN-based methods to higher resolutions by applying morphological transformations and texture changes to reference volumes. Sun et al. \cite{sun2022hierarchical} reached similar resolutions following a hierarchical approach that first generates a low-resolution image and then performs a learned upsampling operation. Liu et al. \cite{liu2023inflating} relied on pretraining 2D models and inflating the 2D convolutions \cite{carreira2017quo} to improve the performance of 3D models.

Dorjsembe et al. \cite{dorjsembe2022three} were the first to adapt the standard Denoising Diffusion Probabilistic Model (DDPM) \cite{ho2020denoising} for modeling 3D medical data, achieving promising results on a brain MR image generation task. Despite their success, the computational complexity and long sampling times associated with simple 3D adaptations posed significant challenges. Peng et al. \cite{peng2023generating} explored a 2.5D approach that models 3D volumes by iteratively predicting 2D slices conditioned on their predecessors and generates new volumes in an autoregressive fashion. Further advances were made by Pinaya et al. \cite{pinaya2022brain} and Khader et al. \cite{khader2023denoising}, who applied latent diffusion models to high-resolution 3D data, effectively reducing the computational complexity and sampling times. However, training the required 3D autoencoder on high-resolution volumes remains a challenging problem and is usually constrained by the available hardware. Friedrich et al. \cite{friedrich2024wdm} proposed applying discrete wavelet transform for spatial dimensionality reduction. This allowed scaling 3D diffusion models to higher resolutions without training an autoencoder and outperformed latent diffusion models on brain MRI and lung CT generation tasks.

\subsection{Image-to-Image Translation}
Multimodal data plays an important role in medical imaging. However, its accessibility is often limited by challenges such as acquisition time and cost, scanner availability, and the risk of additional radiation exposure. To address these limitations, image-to-image translation models aim to generate synthetic images $y$ of a missing modality given an available source modality image $x$. In other words, these models try to find a mapping function $F: X \rightarrow Y$ that maps from the source domain $X$ to the target domain $Y$, such that $y=F(x)$. This problem can be addressed in a paired setting, where training samples $\{x_i, y_i\}_{i=1}^N$ consist of corresponding images $x_i \in X$ and $y_i\in Y$ from the different domains, or in an unpaired setting, where the training data $\{x_i|x_i \in X\}_{i=1}^N$ and $\{y_j|y_j \in Y\}_{j=1}^M$ consists of unrelated samples, requiring different training strategies. Table~\ref{table:overview_i2i} provides an overview of publications on image-to-image translation models for 3D medical images. The table includes information on the translation task, the public datasets used for training, whether the approaches utilized paired or unpaired data, as well as the used network architecture.
\begin{table} 
\begin{center}
\caption{Overview of publications on 3D image-to-image translation frameworks, the translation modalities, public datasets used for training (-- for private data), the type of translated data (paired vs. unpaired), as well as the utilized network architecture.}
\label{table:overview_i2i}
\resizebox{\textwidth}{!}{
\begin{tabular}{ |l|c|p{1.8cm}|p{1.6cm}|c|c|c|c| } 
 \hline
 \textbf{Study} & \textbf{Model} & \textbf{Modality} & \textbf{Dataset(s)} & \textbf{Type} & \textbf{Network(s)}\\ 
 \hline\hline
 Hu et al. (2022) \cite{hu2022domain} & VAE & MRI $\leftrightarrow$ MRI & BRATS & Paired& 2D Spatial-VAE\\
 \hline
 Wei et al. (2019) \cite{wei2019predicting} & GAN & MRI $\rightarrow$ PET & -- & Paired & 3D cGAN \\
 Uzunova et al. (2020) \cite{uzunova2020memory} & GAN & CT $\leftrightarrow$ CT, MRI $\leftrightarrow$ MRI & COPDGene, BRATS & Unpaired & 3D cGAN\\
 Hu et al. (2021) \cite{hu2021bidirectional} & GAN  & MRI $\leftrightarrow$ PET, MRI $\leftrightarrow$ CT & ADNI,  TCIA & Paired & 3D BMGAN\\
 Lan et al. (2021) \cite{lan2021three} & GAN & MRI $\rightarrow$ PET & ADNI & Paired & 3D SC-GAN \\
 Lin et al. (2021) \cite{lin2021bidirectional} & GAN & MRI $\leftrightarrow$ PET & ADNI & Paired & 3D RevGAN\\
 Sikka et al. (2021) \cite{sikka2021mri} & GAN & MRI $\rightarrow$ PET & ADNI & Paired & 3D cGAN \\
 Zhao et al. (2021) \cite{zhao2021mri} & GAN & MRI $\leftrightarrow$ MRI & BRATS & Paired & 3D CycleGAN \\
 Zhang et al. (2022) \cite{zhang2022bpgan} & GAN & MRI $\rightarrow$ PET & ADNI & Paired & 3D BPGAN \\
 Bazangani et al. (2022) \cite{bazangani2022fdg} & GAN & PET $\rightarrow$ MRI & ADNI & Paired & 3D E-GAN\\
 Kalantar et al. (2023) \cite{kalantar2023non} & GAN & CT $\leftrightarrow$ CT & NCCID & Unpaired & 3D CycleGAN\\
 Poonkodi et al. (2023) \cite{poonkodi20233d} & GAN & PET $\leftrightarrow$ CT, PET $\leftrightarrow$ PET, MRI $\leftrightarrow$ MRI & Lung-PET-CT-Dx & Unpaired & 3D CSGAN\\
 Wang et al. (2024) \cite{wang20243d} & GAN & MRI $\rightarrow$ PET & -- & Paired & 3D ViT-GAN\\
 \hline
 Durrer et al. (2023) \cite{durrer2023diffusion} & DDM & MRI $\leftrightarrow$ MRI & -- & Paired & 2D DDPM \\
 Graf et al. (2023) \cite{graf2023denoising} & DDM & MRI $\rightarrow$ CT & MRSpineSeg & Paired & 3D DDIM \\
 Pan et al. (2023) \cite{pan2023cycle} & DDM & MRI $\leftrightarrow$ MRI & BRATS & Paired & 3D DDPM \\
 Zhu et al. (2023) \cite{zhu2023make} & LDM & MRI $\leftrightarrow$ MRI & RIRE & Paired & 2D VAE + 2.5D DDPM/DDIM \\
 Zhu et al. (2024) \cite{zhu2024generative} & LDM & Mask $\rightarrow$ MRI, Mask $\rightarrow$ CT & BRATS, AbdomenCT-1K & Paired & 2D VAE + 2.5D DDIM \\
 Kim et al. (2024) \cite{kim2024adaptive} & LDM & MRI $\leftrightarrow$ MRI & BRATS, IXI & Paired & 3D VQ-GAN + 3D DDPM\\
 Dorjsembe et al. (2024) \cite{dorjsembe2024conditional} & DDM & Mask $\rightarrow$ MRI & BRATS & Paired & 3D DDPM \\
 Pan et al. (2024) \cite{pan2024synthetic} & DDM & MRI $\rightarrow$ CT & -- & Paired & 3D DDPM \\
 Li et al. (2024) \cite{li2024pasta} & DDM & MRI $\rightarrow$ PET & ADNI & Paired & 2.5D DDPM \\
 \hline
\end{tabular}
}
\end{center}
\end{table}

A common task in medical image computing is MRI-to-MRI translation (e.g. ~T1$\leftrightarrow$~T2, or 1.5T~$\leftrightarrow$~3T), which serves several purposes. First, it provides physicians with different contrasts of the images to aid in diagnosis and treatment planning. Second, it can improve the performance of downstream applications such as segmentation tasks by providing the segmentation model with multiple MRI contrasts to work with. Finally, it allows for harmonizing scans from different MR scanners, reducing potential biases in the acquired datasets. These problems have been tackled using VAE-based\cite{hu2021bidirectional}, GAN-based \cite{poonkodi20233d,uzunova2020memory,zhao2021mri} or DDM/LDM-based approaches \cite{durrer2023diffusion,kim20243d,zhu2023make} and have also been addressed in scientific challenges like the MICCAI 2023 Brain MR Image Synthesis for Tumor Segmentation challenge \cite{baltruschat2024brasyn}.

Another application involves generating CT or PET scans from MR images. This enables downstream tasks to be performed on the target modalities without exposing patients to additional CT or PET imaging radiation. For example, Graf et al. \cite{graf2023denoising} performed MRI-to-CT translation to enable segmentation networks trained on CT scans to be applied to MR images.  Recently, the SynthRAD challenge \cite{huijben2024generating}, which aims to provide tools for radiation-free radiotherapy planning by translating MR images to CT scans, has drawn significant attention to this task and highlights the need for well-performing image-to-image translation methods. The task of MRI-to-CT/PET translation has been tackled using different GAN-based \cite{hu2021bidirectional,lan2021three,lin2021bidirectional,sikka2021mri,wang2004image,wei2019predicting,zhang2022bpgan} and DDM-based approaches \cite{li2024pasta,pan2024synthetic}.

Medical image-to-image translation is not only used to translate between different contrasts and modalities but has also been adapted for other tasks, such as anomaly localization, by transforming pathological images into their pseudo-healthy versions \cite{sanchez2022healthy,wolleb2024binary,wolleb2022diffusion}. These approaches, however, have not yet been explored on 3D images.

Image-to-image translation models have proven to be valuable tools for assisting physicians and enabling certain downstream tasks. However, these models have inherent limitations and should not be applied naively. Image-to-image translation models rely on the information present in the input image and the learned prior knowledge of the target modality. This means that if specific clinically relevant details are missing or poorly represented in the source image, they cannot be accurately generated or inferred in the translated image. In addition, these models tend to hallucinate realistic-looking features that do not necessarily correspond to the actual anatomical structures that should be present in the image. As a result, a translated image may contain elements that falsely appear normal or pathological, leading to potential misdiagnosis if naively assumed to be correct.

\subsection{Image Reconstruction}
Reconstructing high-quality images from sparsely sampled or partial measurements is important in speeding up existing medical imaging tools such as CT, PET, or MRI, reducing examination times, harmful radiation exposure to patients, and acquisition costs of these methods. Typical medical image reconstruction tasks include, but are not limited to:
\begin{itemize}
    \item \textbf{Sparse-view computed tomography (SV-CT)}: Aims to reconstruct images from a limited number of X-ray projections.
    \item \textbf{Limited-angle computed tomography (LA-CT)}: Aims to reconstruct images from X-ray projections taken over a limited angular range. Similar to SV-CT, the number of projections is reduced compared to standard CT, but the acquisition angle is additionally limited, e.g. by physical constraints in the operating room.
    \item \textbf{Low-dose CT denoising (LDCT-D)}: Aims to remove noise from low-dose CT scans, enhancing image quality to a level comparable to standard-dose CT scans.
    \item \textbf{Compressed-sensing magnetic resonance imaging (CS-MRI)}: Aims to reconstruct high-quality images from undersampled MRI data.
    \item \textbf{Z-axis super-resolution on MR images (ZSR-MRI)}: Aims to enhance the resolution of MR images along the z-axis.
    \item \textbf{Standard-dose PET (SPET) scans from Low-dose PET (LPET) scans}: Aims to reconstruct high-quality PET images from low-dose PET scans.  
\end{itemize}
While these tasks have extensively been explored on 2D images (sliced volumes), research on directly solving these problems on the 3D data is limited. An overview of publications on 3D medical image reconstruction is shown in Table~\ref{table:overview_rec}.
\begin{table} 
\begin{center}
\caption{Overview of publications on 3D image reconstruction frameworks, the data modalities, public datasets used for training(-- for private data), the solved tasks, as well as the utilized network architecture.}
\label{table:overview_rec}
\resizebox{\textwidth}{!}{
\begin{tabular}{ |l|c|c|p{2.8cm}|p{2.4cm}|c| } 
 \hline
 \textbf{Study} & \textbf{Model} & \textbf{Modality} & \textbf{Dataset(s)} & \textbf{Task} & \textbf{Network(s)}\\ 
 \hline\hline
 Wolterink et al. (2017) \cite{wolterink2017generative} & GAN & CT & -- & LDCT-D & 3D GAN\\
 Wang et al. (2018) \cite{wang20183d} & GAN & PET & -- & LPET-SPET & 3D cGAN\\
 Luo et al. (2021) \cite{luo20213d} & GAN & PET & -- & LPET-SPET & 3D ViT-GAN\\
 Zeng et al. (2022) \cite{zeng20223d} & GAN & PET & -- & LPET-SPET & 3D ViT-GAN\\
 Xue et al. (2023) \cite{xue2023cg} & GAN & PET & Ultra-low Dose PET Imaging Challenge 2022 & LPET-SPET & 3D SRGAN\\
 Wang et al. (2024) \cite{wang20243d} & GAN & PET & -- & LPET-SPET & 3D ViT-GAN\\
 \hline
Lee et al. (2023) \cite{lee2023improving} & DDM & CT, MRI & AAPM 2016 CT Low-Dose
Grand Challenge & SV-CT, CS-MRI, ZSR-MRI & 2D Score DDMs\\
Chung et al. (2023) \cite{chung2023solving} & DDM & CT, MRI & AAPM 2016 CT Low-Dose
Grand Challenge, BRATS, fastMRI & SV-CT, LA-CT, CS-MRI & 2D Score DDM\\
Xie et al. (2023) \cite{xie2023dose} & DDM & PET & -- & LPET-SPET & 2.5D DDPM/DDIM\\
He et al. (2024) \cite{he2024blaze3dm} & DDM & CT, MRI &  AAPM 2016 CT Low-Dose
Grand Challenge, IXI & SV-CT, LA-CT, CS-MRI, ZSR-MRI & Triplane DDPM\\
Li et al. (2024) \cite{li2024two} & DDM & CT, MRI & AAPM 2016 CT Low-Dose
Grand Challenge, BRATS & SV-CT, CS-MRI & 2D Score DDMs \\
 \hline
\end{tabular}
}
\end{center}
\end{table}

Several scientific challenges have drawn attention to the topic and provided valuable datasets for evaluating different image reconstruction approaches. These challenges include the AAPM 2016 CT Low-Dose Grand Challenge \cite{mccollough2017low}, the MICCAI 2021 Brain MRI Reconstruction Challenge with Realistic Noise \cite{melanie_ganz_2021_4572640}, the MICCAI 2022 Ultra-low Dose PET Imaging Challenge \cite{kuangyu_shi_2022_6361846}, and the MICCAI 2023 Cardiac MRI Reconstruction Challenge \cite{chengyan_wang_2023_7840229}. Due to the computational complexity of directly handling 3D data, most presented approaches still operate on 2D slices, highlighting the need for efficient 3D backbones. Existing 3D GAN-based approaches have mostly focused on synthetic SPET image generation with conditional GAN \cite{wang20183d}, Vision Transfromer GANs (ViT-GAN) \cite{luo20213d,wang20243d,zeng20223d} or Classification-Guided GAN \cite{xue2023cg} and operate on rather low-resolution volumes. Diffusion-based approaches primarily focused on SV-CT, LA-CT and CS-MRI. They formulate the task as an inverse problem of predicting an unknown image $x$ given limited measurements $y$. The forward model is denoted as $y = \mathbf{A}x+\epsilon$, with $\mathbf{A}$ being a degradation function (e.g. partial sampling of the sinogram for SV-CT and LA-CT, or $k$-space for CS-MRI) and the measuring noise $\epsilon$. The inverse task $\hat{x}=G_\theta(y)$ is solved using a DGM $G_\theta$. They address the problem of dealing with high-dimensional 3D data by applying perpendicular 2D models \cite{lee2023improving,li2024two}, by conditioning the model on adjacent slices to form 2.5D models\cite{xie2023dose}, or by applying 2D models to a triplane representation of the data \cite{he2024blaze3dm}. 

Similar to image-to-image translation models, image reconstruction models can hallucinate structures that are not present in the actual anatomy, potentially leading to misdiagnosis. To ensure clinically relevant and reliable images, it is essential to develop robust models, compile comprehensive training datasets, and perform rigorous validation. In addition, these models should be used with caution and an understanding of the potential risks.

\section{Evaluating Deep Generative Models}
\label{sec:eval}
Evaluating deep generative models in medical imaging is not trivial and probably deserves its own chapter. Nevertheless, we will give a brief overview of popular metrics and discuss various image quality, diversity, utility, privacy, and other non-image-related metrics that should be considered when evaluating generative models and the data they synthesize.
\subsection{Image Quality Metrics}
\label{subsec:qual_metrics}
The \textbf{Fréchet Inception Distance (FID)} \cite{heusel2017gans} is a metric that measures image fidelity by comparing the distribution of real and generated images without requiring image pairs. It has widely been applied to evaluate unconditional image-generation tasks. The FID score is calculated by first extracting high-level features from real and synthetic images using an intermediate activation of a pre-trained neural network, calculating statistics over these features by fitting two multivariate Gaussians to the real $\mathcal{N}(\mu_{r}, \Sigma_{r})$ and synthetic images features $\mathcal{N}(\mu_{s}, \Sigma_{s})$, and computing the Fréchet distance between those distributions:
\begin{equation}
    FID = \|\mu_{r} - \mu_{s}\|_2^2 + tr(\Sigma_{r} + \Sigma_{s} - 2\sqrt{\Sigma_{r}\Sigma_{s}}).
\end{equation}
A small FID score indicates that the distributions of real and synthetic images are similar, suggesting that the model effectively learned the data distribution, which results in a good visual appearance. In 3D medical image computing, Med3D-Net \cite{chen2019med3d} has widely been applied as a feature extraction network. While FID scores are a useful tool for assessing image fidelity in unconditional image generation tasks, comparing these scores across publications is not straightforward and should be done cautiously. This is because FID scores are highly dependent on the choice of feature extraction network and the specific feature layer used. Further, metrics like FID, or variants like the Kernel Inception Distance (KID) \cite{binkowski2018demystifying}, are highly dependent on a sufficiently large number of samples used to approximate the data distributions \cite{chong2020effectively}. Without fulfilling this requirement, these metrics lose their meaningfulness, which is an often overlooked problem within limited data regimes such as medical imaging.

For image generation tasks where a ground truth image is available, such as a paired cross-modality image synthesis task, metrics like Peak Signal-to-Noise-Ratio (PSNR), Structural Similarity Index Measure (SSIM) or Mean Squared Error (MSE) can be applied to compare generated and ground truth image. The \textbf{Peak Signal-to-Noise-Ratio (PSNR)} is a metric originally used to measure the reconstruction quality of lossy compressed images. In the context of evaluating conditional image generation, it measures the fidelity of the synthetic image $S$ compared to the real ground truth $R$ and is defined as
\begin{equation}
    PSNR = 10\cdot\log_{10}\left(\frac{{MAX}_R^2}{MSE}\right) = 20\cdot\log_{10}({MAX}_R)-10\cdot\log_{10}(MSE),
\end{equation}
with $MAX_R$ being the maximum possible intensity value of the images, usually $255$ for uint8 grayscale images, and the \textbf{Mean Squared Error (MSE)} between real and synthetic image, which is defined over all voxels $N$ of the volume:
\begin{equation}
    MSE = \frac{1}{N}\sum_{i=1}^{N}(R_{i} - S_{i})^{2}.
\end{equation}

\begin{figure}
    \centering
    \includegraphics[width=.95\textwidth]{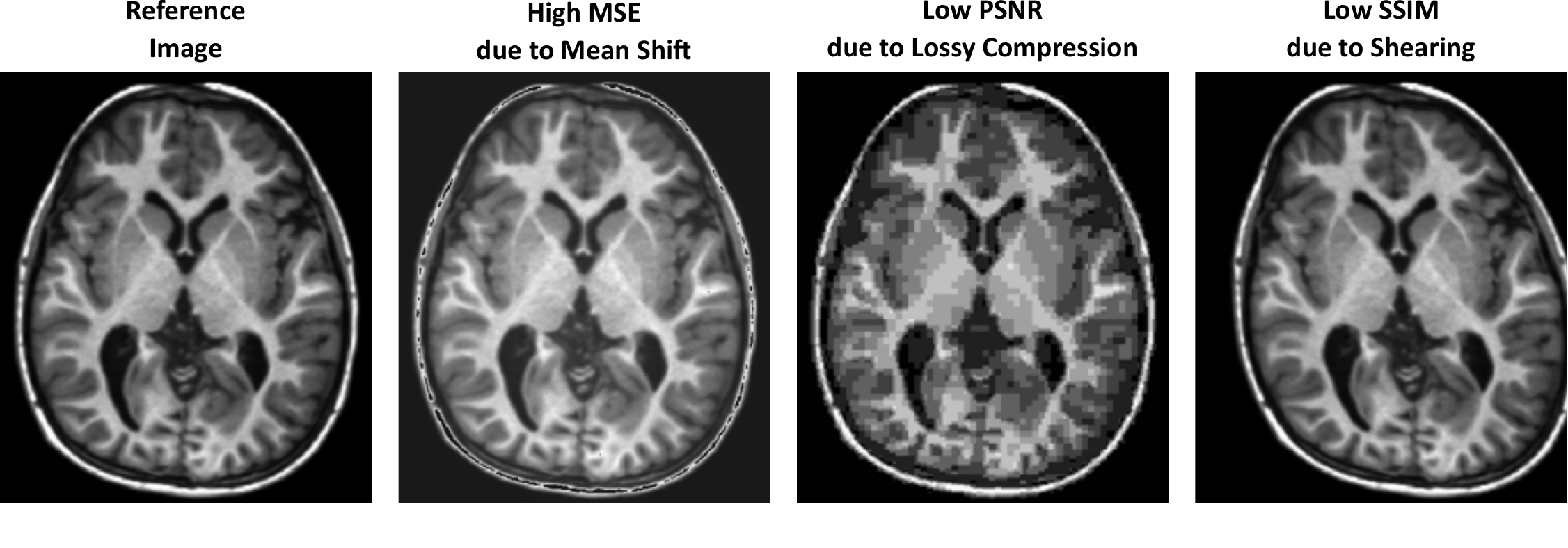}
    \caption{Paired Image Quality Metrics. Paired Metrics allow quantifying the similarity of synthetic or corrupted images to a reference image (first column). Examples include the \emph{Mean Squared Error (MSE)}, commonly used to quantify noise artifacts (second column), the \emph{Peak Signal-to-Noise-Ratio (PSRN)}, often used to quantify compression quality (third column), as well as the \emph{Structural Similarity Index Measure (SSIM)} to evaluate structural distortions (fourth column).}
    \label{fig:metrics_overview}
\end{figure}

While PSNR and MSE are widely applied metrics for assessing image quality, they have several limitations. As pixel-level metrics, they can't capture any structural information that is strongly correlated with good visual perception of the generated images. \cite{wang2004image}
This drawback led to the development of the \textbf{Structural Similarity Index Measure (SSIM)}, which evaluates the perceived change in structural information, luminance, and contrast. It is computed over a shifting window of similar size, e.g. $11\times 11\times 11$, denoted as $r$ for a window of $R$ and $s$ for the same window from $S$. The SSIM is defined as:
\begin{equation}
    SSIM = \frac{{(2\mu_r \mu_s + c_1)(2\sigma_{rs} + c_2)}}{{(\mu_r^2 + \mu_s^2 + c_1)(\sigma_r^2 + \sigma_s^2 + c_2)}}
\end{equation}
with the mean $\mu_r$ and $\mu_s$, and variances $\sigma_r^2$ and $\sigma_s^2$ over the respective windows pixel intensities, the covariance $\sigma_{rs}$ between them, as well as two constants $c_1$ and $c_2$ to numerically stabilize the score against division with weak denominators. While a small SSIM score indicates that two images significantly differ in structural information, luminance and contrast, a score close to 1 indicates high image similarity. Example image variations corresponding to significant changes in MSE, PSNR and SSIM scores are visualized in Figure~\ref{fig:metrics_overview}. In addition to the classic SSIM score, several variants such as MS-SSIM \cite{wang2003multiscale}, or 4-(G-)SSIM \cite{li2010content} have been proposed to improve the quality metric further. A competitive evaluation of several SSIM versions on radiology images has been performed in \cite{renieblas2017structural}, suggesting that 4-MS-G-SSIM provides optimal results and strongly agrees with human perception.
Other ways to assess image quality are the \textbf{Inception Score} \cite{salimans2016improved}, \textbf{Precision} \cite{kynkaanniemi2019improved,sajjadi2018assessing}, or by conduction a \textbf{Visual Turing Test} \cite{geman2015visual}.
\subsection{Image Diversity Metrics}
\label{subsec:div_metrics}
While the \textbf{Structural Similarity Index Measure (SSIM)} can be used to measure the similarity between image pairs, it has also been applied to assess the diversity of the generated images in unconditional generation tasks, an important factor to consider when evaluating the performance of generative models. In \cite{friedrich2024wdm} and \cite{pinaya2022brain}, the authors measure image diversity by averaging the MS-SSIM over generated images by iteratively comparing a reference image to all other generated images. A low SSIM in this case indicates high generation diversity, meaning that the generated images are not similar to each other. Such quantification of diversity can also be carried out in the feature space of pre-trained feature extractors, using similarity metrics like LPIPS \cite{zhang2018unreasonable}.

Another common way to assess image diversity is the \textbf{Recall Score} \cite{kynkaanniemi2019improved,sajjadi2018assessing}, which measures the fraction of the training data manifold that can be produced by the generative network. Similar to computing the FID score, a feature extraction network is applied to produce a set of high-level features of real $\Phi_r$ and synthetic images $\Phi_s$. A feature vector of a single image is denoted as $\phi_r$ for a real image and as $\phi_s$ for a synthetic image. The Recall score is then defined as
\begin{equation}
    \text{Recall} = \frac{1}{|\Phi_r|} \sum_{\phi_r \in \Phi_r}f(\phi_r, \Phi_s)
\end{equation}
with $|\Phi_r|$ being the number of images to compute the score, and $f(\phi_r, \Phi_s)$ a nearest-neighbor-based binary function that determines whether a real image could be generated by evaluating whether it lays within the approximated synthetic data distribution. A high recall score indicates that the trained network generates diverse samples from the entire data distribution, while a low recall score could be a sign of mode collapse.
\subsection{Utility \& Privacy Metrics}
\label{subsec:ut_and_priv_metrics}
Another common way to evaluate the performance of deep generative networks is to assess the utility of the generated data by measuring \textbf{performance improvements on relevant downstream tasks} when trained with additional synthetic data. In medical image computing, this has, for example, been done by measuring performance improvements of classification \cite{frid2018gan,frisch2023synthesising,sagers2023augmenting} or segmentation \cite{al2023usability} tasks.

Since deep generative models are known to memorize data \cite{van2021memorization,dar2024unconditional,gu2023memorization}, assessing the privacy of synthetic images and minimizing the risk of re-identification \cite{fernandez2023privacy,yoon2020anonymization} before sharing images or weights of models trained on non-public data is crucial. To evaluate the privacy of generated images, metrics such as the \textbf{Rarity Score} \cite{han2022rarity} or \textbf{Average Minimum Cosine Distance (AMD)} \cite{bai2021training}, which measures the uncommonness of generated images as the nearest-neighbor distance of real and synthetic data points in a latent space, have been proposed. Other privacy metrics rely on \textbf{model-based re-identification} \cite{packhauser2022deep} or apply \textbf{Extraction Attacks} \cite{carlini2023extracting} to measure privacy by trying to extract training images from the trained models. 
\subsection{Non-Image Metrics}
\label{subsec:non-img_metrics}
In addition to the previously discussed image metrics, other factors such as the model's \textbf{Inference Time}, \textbf{Computational Efficiency}, usually measured in Floating Point Operations (FLOPs), or \textbf{Memory Consumption} must also be considered in the evaluation to ensure the model's applicability to real-world problems. This is especially important for time-critical applications or when the method is to be used in a resource-constrained environment. The \textbf{Ethical and Social Impact} of the model, i.e. possible bias and fairness of the model as well as potential misuse, should also be considered.

\section{Current Challenges \& Conclusion}
\label{sec:chall&outlook}
Despite the recent success of deep generative models in medical imaging, many essential and critical challenges remain. This section overviews such open challenges and potential directions for future research.

\subsection{Challenges}
High-resolution data is crucial for a thorough and accurate analysis of medical images. However, many current deep generative models struggle with scaling to such high-dimensional spaces. Additionally, computational resources are often limited in clinical settings, making in-house training of large-scale generative models nearly impossible. Furthermore, medical image analysis often benefits from longitudinal data, e.g. repeated scans of patients at progressive points in time. Modeling such temporal data with generative models further increases the required compute significantly and is an open and highly relevant research direction \cite{puglisi2024enhancing,stolt2023nisf,yoon2023sadm,zhu2024loci}. To address these challenges, various methods have been proposed to significantly reduce the number of model parameters, speed up training and inference times, and lower GPU memory requirements. Notable approaches include WDMs \cite{friedrich2024wdm,phung2023wavelet} and Neural Cellular Automata \cite{kalkhof2024frequency}. Despite these advancements, the \textbf{need for scalable and efficient models} continues to drive further research in this area.

The scalability problem also applies to the data itself. Many state-of-the-art generative models are trained on excessive amounts of data that are typically unavailable for most medical problems. Collecting and annotating such data can be a resource- and labor-intensive process, further complicated by regulatory issues. In addition, the datasets used to train generative models should be as diverse and representative as possible to avoid the negative effects of bias. One way to mitigate these problems and allow for the development of a robust and useful model is to \textbf{provide data in an open access paradigm}. Another way to effectively train generative models and fully capture complex data distributions is to \textbf{develop efficient methods for federated learning} \cite{de2024training,jothiraj2023phoenix,tun2023federated} and solve existing problems related to this. An example of such issues could be the case where a patient - or even an entire institute - opts out of a study and withdraws its samples from the training distribution. The process of unlearning \cite{gandikota2023erasing}, therefore, needs to be addressed in the context of generative models for medical imaging.

Another commonly faced challenge is related to the evaluation of DGMs in the medical domain. Most metrics widely used to evaluate generative models in the natural image domain must only be carefully applied to medical images. These metrics might only provide reliable quantitative results with adequately pre-trained feature extractors \cite{barratt2018note} and an appropriately large test sample size \cite{chong2020effectively}. In addition, the three-dimensionality of medical imaging data is often not considered in such metrics, e.g. feature extractors for 3D data might not be publicly available and have to be built from scratch. We, therefore, identify the \textbf{need for more rigorous quantitative testing of generative models in the medical domain}. 

Further challenges arise when it comes to safely deploying these algorithms, as generative models can pose the threat of (unconscious or deliberate) data corruption \cite{hataya2023will}. Especially when these models are deployed as a data augmentation method for downstream task models, additional curation is needed to minimize the risk of error propagation from the generative to the downstream task model \cite{van2023synthetic}. Additionally, the risk of these models hallucinating can potentially lead to drastic consequences, e.g. for the tasks of reconstruction or inpainting in the medical domain. 

Most generative models in the medical domain are developed and trained on publicly available data. In some cases, in which private data cohorts are also considered, privacy protection is an important issue that needs to be taken into account. Even releasing the weights of generative models trained on private data could lead to privacy violations, since the training data can essentially be extracted from these models \cite{carlini2023extracting}. The problem of \textbf{privacy preserving generative models} needs to be addressed and is an interesting and promising direction for future research.

\subsection{Conclusion}
Deep generative models have achieved significant success in recent years, proving to be a valuable tool for learning complex data distributions and solving medically relevant downstream tasks. We reviewed the background of Variational Autoencoders, Generative Adversarial Networks, and Denoising Diffusion Models, highlighting their strengths and weaknesses. In addition, we demonstrated their application to tasks such as unconditional image generation, image-to-image translation, and image reconstruction, and discussed commonly used evaluation metrics as well as pitfalls associated with these metrics.

Despite the mentioned advances in DGMs for 3D medical image synthesis, several open challenges remain that motivate future research in this area. Finding novel, more efficient data representations, developing tools for federated learning, or exploring methods to address unlearning and privacy concerns are critical to advancing the capabilities of deep generative models in the medical domain. The goal of these efforts is to create efficient, fair, and reliable algorithms that can provide physicians with valuable insights and ultimately improve personalized medicine, enhance predictive analytics, and facilitate the development of new therapeutic strategies.

%
%
\begin{acknowledgement}
This work was financially supported by the Werner Siemens Foundation through the MIRACLE II project.
\end{acknowledgement}
%
%
\ethics{Competing Interests}{
The authors have no conflicts of interest to declare that are relevant to the content of this chapter.}

\eject
\bibliographystyle{styles/spmpsci.bst}
\bibliography{bibliography}

\end{document}